\documentclass[12pt]{article}
\usepackage{amsmath}
\usepackage{amsfonts}
\usepackage{amssymb}
\usepackage{mathrsfs}
\usepackage{multicol}
\usepackage{color}
\usepackage{graphicx}
\usepackage{pstricks}
\usepackage{pst-node}
\usepackage{wrapfig}
\usepackage{enumerate}
\usepackage{caption}
\usepackage{float}
\usepackage{pdfpages}
\usepackage{cite}
\usepackage{subfigure}
\usepackage{caption}
\usepackage{amsfonts,amssymb}
\usepackage{amsthm}
\usepackage{amscd}
\usepackage{caption}
\usepackage{wrapfig}

\definecolor{gray1}{gray}{0.1}
\definecolor{gray2}{gray}{0.2}
\definecolor{gray3}{gray}{0.3}
\definecolor{gray4}{gray}{0.4}
\definecolor{gray5}{gray}{0.5}
\definecolor{gray6}{gray}{0.6}
\definecolor{gray7}{gray}{0.7}
\definecolor{gray8}{gray}{0.8}
\definecolor{gray9}{gray}{0.9}

\newpsobject{showgrid}{psgrid}{subgriddiv=1,griddots=10,gridlabels=6pt}

\definecolor{dark-green}{rgb}{0,0.7,0}
\definecolor{dark-blue}{rgb}{0,0.2,0.5}
\definecolor{med-blue}{rgb}{0,0.7,1}
\definecolor{mblue}{rgb}{0,0.2,1}
\definecolor{cnc}{rgb}{0.8,0,0}
\definecolor{light-red}{rgb}{1,0.8,0.8}
\definecolor{dark-yelow}{rgb}{1,0.8,0}
\definecolor{light-blue}{rgb}{0.8,0.9,1}
\definecolor{verylight-blue}{rgb}{0.93,0.95,1}
\definecolor{light-yelow}{rgb}{1,0.9,0.8}
\definecolor{grey}{gray}{0.88}

\newpsobject{showgrid}{psgrid}{subgriddiv=1,griddots=10,gridlabels=6pt}

\newcommand{\be}{\begin{equation}}
\newcommand{\ee}{\end{equation}}
\newcommand{\bea}{\begin{eqnarray}}
\newcommand{\eea}{\end{eqnarray}}
\newcommand{\beann}{\begin{eqnarray*}}
\newcommand{\eeann}{\end{eqnarray*}}
\def\IZ{\rlx\hbox{\sf Z\kern-.4em Z}}
\def\IR{\rlx\hbox{\rm I\kern-.18em R}}
\def\IC{\rlx\hbox{\,$\inbar\kern-.3em{\rm C}$}}
\def\one{\hbox{{1}\kern-.25em\hbox{l}}}

\begin{document}

\thispagestyle{empty}

\setlength{\abovecaptionskip}{10pt}

\begin{center}
{\Large\bfseries\sffamily{A zero-curvature representation of electromagnetism and the conservation of electric charge}}
\end{center}
\vskip 3mm
\begin{center}
{\bfseries{\sffamily{G. Luchini$^{\rm 1}$, V. B. Zaché$^{\rm 1}$}}}\\
\vskip 3mm{
$^{\rm 1}$\normalsize Departamento de F\'isica,
Universidade Federal do Esp\'irito Santo (UFES),\\
CEP 29075-900, Vit\'oria-ES, Brazil}
\end{center}

\begin{abstract} 
We show that the laws of electromagnetism in $(D+1)$-dimensional Minkowski space-time $\mathcal{M}$, explicitly for $D=1$, $2$ and $3$, can be obtained from an integral representation of the zero-curvature equation in the corresponding loop space $\mathcal{L}^{(D-1)}(\mathcal{M})$. The conservation of the electric charge can be seen as the result of a hidden symmetry in this representation of the dynamical equations.
\end{abstract}

\section{Introduction}

Hidden symmetries play a special role in the construction of soliton solutions and their conserved charges in integrable field theories\cite{babelon}. In $(1+1)$-dimensional space-time these theories will generally admit what is called a zero-curvature representation\cite{lax,fadeev} in which their dynamical equations become equivalent to the flatness of a Lie algebra-valued 1-form connection whose components are functions of the physical fields and their derivatives. 

This representation lies in the core of the construction of conserved charges\cite{laf_charges} which are obtained from the holonomy operator associated to that 1-form connection: its flatness implies the path-independence of the holonomy in space-time and leads to an isospectral evolution of the holonomy evaluated over space; the preserved eigenvalues of this operator are the conserved charges.

In integrable field theories, some of these charges may coincide with those obtained from Noether's theorem but the symmetries here are hidden in the gauge invariance of the zero-curvature equation, which also defines the ground for the development of algebraic methods to construct soliton solutions. 

The possibility of extending the zero-curvature formulation to field theories in higher-dimensional space-time as an attempt to understand integrability in this context was explored in \cite{laf_1997}. The crucial step towards this approach was the generalization of the holonomy operator through a non-abelian Stokes theorem and the interpretaion of this construction in loop space\cite{laf_2010}. In \cite{luchini1,luchini2} it was shown how this non-abelian Stokes theorem can be used to define the integral version of Yang-Mills equations leading to dynamically gauge-invariant conserved charges. Some consequences of this formulation of gauge theories in loop space were discussed in \cite{luchini3, luchini4}.

In the present paper we show that the integral equations of electromagnetism can be represented in loop space by an integral equation for a flat connection $\mathcal{A}$. In a Minkowski $(D+1)$-dimensional space-time $\mathcal{M}$, this connection is constructed in the corresponding loop space which is defined by the maps $\mathcal{L}^{(D-1)}(\mathcal{M}) = \left\{\Gamma: S^{D-1}\to \mathcal{M} \vert \Gamma(0) = x_R \right\}$, taking $(D-1)$-dimensional spheres in space-time, based at $x_R$, to points in this loop space. The connection is defined in terms of an exact $D$-form in space-time evaluated on the loop $S^{D-1}$. The flatness of the loop space connection, $\delta \mathcal{A} = 0$, will give the local conservation of the electric charge and the charge itself can be obtained as the eigenvalues of the generalized $D$-holonomy evaluated over the space as a consequence of the (hidden) symmetry of the integral equation in loop space: its invariance under homotopic transformations of the path. 

The integral equation in loop space is written in space-time as a Lorentz scalar integral equation and by choosing appropriately a space-time slicing with $D$-dimensional hyper-volumes one can recover the usual integral expressions, namely Gauss', Faraday and Amp\`ere-Maxwell laws.

\section{The integral equations of electromagnetism in $1+1$ dimensions}

The Maxwell equations in 2-dimensional space-time, in Gaussian coordinates, are given by\footnote{The coordinates of the $(D+1)$-dimensional Minkowski space-time are $x^\mu = (ct,x^i)$, $\mu=0,1,\dots,D$ and the Minkowski metric has signature $\eta_{\mu\nu}=\textrm{diag}(1,-1,\dots,-1)$.}
\begin{equation}
\label{eq: max_1d}
    \partial_\mu F^{\mu \nu} = \frac{2}{c} J^\nu \quad \mu ,\nu \ = 0, 1 
\end{equation}
where $J^\mu = (c\rho , j)$ is the covariant electric current density and the Faraday tensor defined in terms of the gauge potential $a_\mu$ is given by $F_{\mu\nu} = \partial_\mu a_\nu - \partial_\nu a_\mu$ having $F_{01} = E = -F_{10}$ the only non-vanishing components, with $E$ the electric field.

We consider the loop space $\mathcal{L}^{(0)}(\mathcal{M})$. This is the space of maps $\Gamma$ from the sphere $S^0$ into the space-time manifold $\mathcal{M}$ defined as $\mathcal{L}^{(0)}(\mathcal{M}) = \left\{\Gamma: S^{0}\to \mathcal{M} \vert \Gamma(0) = x_R\right\}$ where $x_R$ is a fixed point in $\mathcal{M}$ which we call the reference point. The image of the map $\Gamma$ will be, in this case, also points in the loop space $\mathcal{L}^{(0)}(\mathcal{M})$. 

Let us define a $1$-form connection $\mathcal{A}$ in this loop space\cite{laf_2010} as
\begin{align}
    \mathcal{A}=A_\mu \delta x^\mu.
\end{align}

We want to find an integral representation of the Maxwell equations in $\mathcal{L}^{(0)}(\mathcal{M})$ based on a flat connection in this space. This can be done if we write the field in space-time $A_\mu$ as the components of the exact $1$-form $A=df = \partial_\mu f dx^\mu$, since then $\delta \mathcal{A} =  \frac{1}{2}\left(\partial_\mu A_\nu - \partial_\nu A_\mu \right)\delta x^\mu \wedge \delta x^\nu = 0$.

Let us write $f$ in terms of the physical fields as
\begin{align}
    f=\frac{i\beta}{2}\epsilon^{\mu \nu}F_{\mu\nu},
\end{align}
where $\beta$ is an arbitrary constant.

Using the equations of motion we obtain the components $A_\mu$ as
\begin{align}
    A_\mu = \frac{i2\beta}{c}\epsilon_{\mu\nu}J^\nu
\end{align}
and consequently the loop space connection is given by
\begin{align}
    \mathcal{A} = \frac{i2\beta}{c}\epsilon_{\mu\nu}J^{\nu} \delta x^{\mu}.
\end{align}

Given the flatness of the connection we can associate its integral over a path $\Gamma$ in loop space with the value of a potential $\varphi$ evaluated at the borders of this path. This is the integral representation of the zero-curvature equation for the connection in loop space:
\begin{align}
    \Delta \varphi = \int_{\Gamma} \mathcal{A}(\sigma ) \ d\sigma
    \label{eq: loopspace_1d}
\end{align}
where $\mathcal{A}(\sigma) \equiv A_\mu \frac{dx^{\mu}}{d\sigma}d\sigma$, with $\sigma$ parameterizing the path $\Gamma$. Clearly, by construction we have that $\varphi = f$ and $A_\mu = \partial_\mu f$. 

From the definitions of $f$ and $A_\mu$ in terms of the physical fields and their local relations given by equation (\ref{eq: max_1d}) this equation becomes
\begin{align}
    E(x) - E(x_R) = \frac{2}{c}\int_0 ^{2\pi} \epsilon_{\mu \nu}J^{\nu}\frac{dx^\mu}{d\sigma}d\sigma.
    \label{eq: integral_1d}
\end{align}

\begin{figure}[h]
\centering
\includegraphics[scale=0.5]{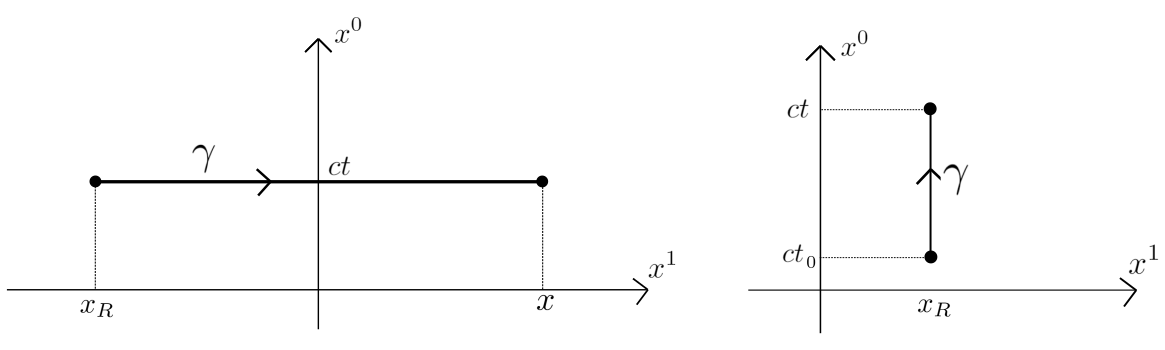}
\caption{With an appropriate choice for the paths in space-time the Lorentz scalar integral equations will give the usual integral expressions for the laws of electrodynamics.}
\label{fig56}
\end{figure}

This is the Lorentz scalar integral equation of electromagnetism in $1+1$ dimensional space-time. In order to obtain the usual version of the integral equations, here equivalent to the Gauss' and Amp\`ere's laws for the electric field, we need to specify the curve $\gamma$ is space-time where we integrate the dual of the electric current (see figure \ref{fig56}).

When $\gamma$ is considered to be purely spatial at a constant time $t$ we have, from (\ref{eq: integral_1d})
\begin{equation}
\label{eq: gauss_1d}
     E(t,x) - E(t,x_R) = \frac{2}{c}\int_{x_R} ^{x} \epsilon_{1 0}J^{0}\frac{dx^1}{d\sigma}d\sigma = 2\int_{x_R} ^{x}\rho dx'
\end{equation}
which can be recognized as the Gauss law for the electric field. 

Next, the Ampère-Maxwell law follows from the integral equation (\ref{eq: integral_1d}) when we consider the curve $\gamma$ to have constant spatial coordinate $x$:
\begin{equation}
\label{eq: ampere_1d}
    E(t,x) - E(0,x) = \frac{2}{c}\int_{0} ^{t} \epsilon_{0 1}J^{1}\frac{dx^0}{d\sigma}d\sigma 
= -2\int_{0} ^{t} j dt'.
\end{equation}

We notice that if the curve $\gamma$ is taken to be infinitesimal, i.e., $x = x_R + \delta x$, then the differential equations (\ref{eq: max_1d}) can be recovered from (\ref{eq: gauss_1d}) and (\ref{eq: ampere_1d}):
\begin{align}
    &E(t,x) - E(t,x_R) \approx 2\rho \Delta x \quad \Longrightarrow \quad \frac{\partial E}{\partial x} = 2\rho \nonumber \\
    &E(t,x) - E(0,x) \approx -2j \Delta t \quad \Longrightarrow \quad \frac{\partial E}{\partial t} = -2j. \nonumber
\end{align}

\subsection{The conservation of the electric charge in 2 dimensions}

The loop space representation of the integral dynamical equations defines a relation between the 0-form $\varphi$ at the borders of the path $\Gamma$ with the integral of the 1-form connection $\mathcal{A}$ along this path. 

Considering an infinitesimal variation of $\Gamma$ keeping its borders fixed, we find from (\ref{eq: loopspace_1d}) that

\begin{eqnarray*}
    \delta \left(\Delta \varphi\right) = 
    \int_{\Gamma} (\partial_\nu A_\mu - \partial_\mu A_\nu ) \frac{dx^{\mu}}{d\sigma} \delta x^{\nu} d\sigma. 
\end{eqnarray*}
Since the border of $\Gamma$ remains fixed, $\Delta \varphi$ should not change and consequently the l.h.s of the equation above vanishes. On the other hand, given that $A$ is exact, also the r.h.s above vanishes and we conclude that equation (\ref{eq: loopspace_1d}) remains invariant under homotopic deformations of the path $\Gamma$. In other words, $\varphi$ is path-independent and equivalently $\mathcal{A}$ is flat. Writing the loop space connection in terms of the physical fields we have
\begin{equation}
    \delta \mathcal{A} = \partial_\nu A_\mu \delta x^{\nu} \land \delta x^{\mu}
    =\frac{i2\beta}{c}\epsilon^{\nu\mu}\epsilon_{\mu\lambda}\partial_\nu J^{\lambda}\delta x^0\delta x^1 =\frac{i4\beta}{c}\partial_\lambda J^{\lambda}\delta x^0\delta x^1
\end{equation}
and consequently $ \delta \mathcal{A}=0$ implies $\partial_\mu J^\mu =0$, i.e, the local conservation of the electric charge is obtained as a consequence of the zero-curvature of the loop space connection.

This path-independence of the integral equation (\ref{eq: loopspace_1d}) in loop space is the hidden symmetry behind this conservation law . 

The conserved charges associated to this symmetry can be obtained from the holonomy operator defined by the parallel transport equation along a curve $\gamma$ in space-time parameterized by $\sigma \in [0,2\pi]$,
\begin{align}
    \frac{dW}{d\sigma} +A_\mu \frac{dx^\mu}{d\sigma}W =0,
\end{align}
whose solution can be formally written as 
\begin{align}
\label{eq: wsol}
    W_\gamma = e^{-\int_0 ^{2\pi}A_\mu \frac{dx^\mu}{d\sigma}d\sigma}W_\circ, 
\end{align}
where $W_0$ is obtained from the initial conditions. 

\begin{figure}[H]
    \centering
    \includegraphics[scale = 0.4]{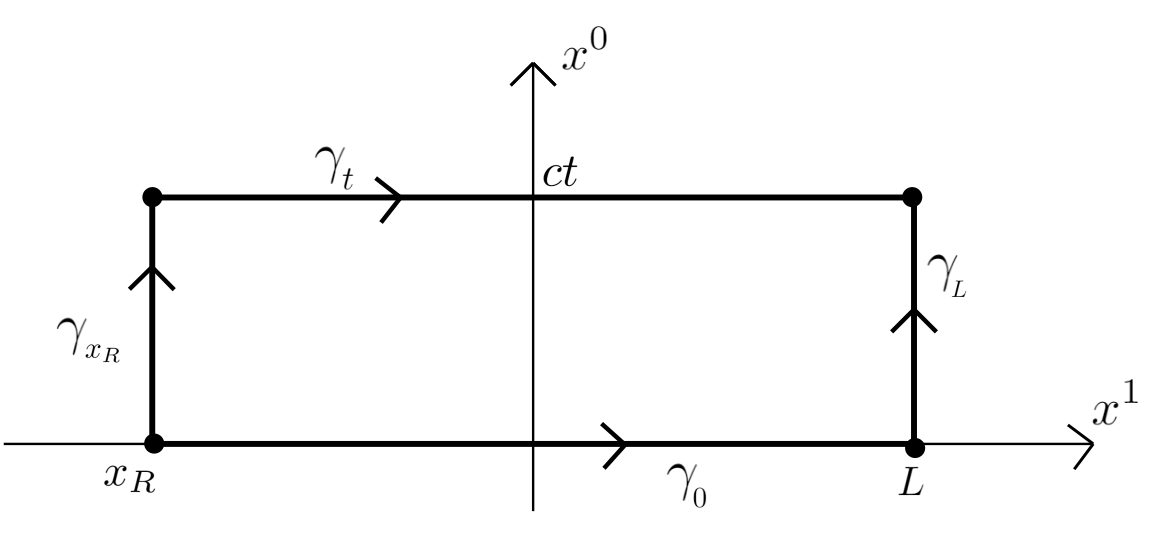}
    \caption{The curve in the $(1+1)$-dimensional space-time is equivalent to the path defined in the corresponding loop space.}
    \label{fig1}
\end{figure}

We consider the paths given in figure \ref{fig1} joinning the points $x_R$ and $x = (ct,L)$. We assume that the two paths $\gamma_L \circ \gamma_0$ and $\gamma_t \circ \gamma_{x_R}$ can be deformed into each other by continuous transformations $x^{\mu} \to x^{\mu}+\delta x^{\mu}$ which make the holonomy $W$ calculated over $\gamma_L \circ \gamma_0$, changes as\footnote{See appendix \ref{app}}
\begin{equation}
\label{eq: varw}
    \delta W = \int_0 ^{2\pi} (\partial_\mu A_\nu -\partial_\nu A_\mu ) \frac{dx^{\mu}}{d\sigma} \delta x^{\nu}d\sigma.
\end{equation}

The r.h.s of equation (\ref{eq: varw}) vanishes since $A$ is an exact $1$-form, giving $\delta W =0$, which means that the holonomy is in fact path-independent. So, the holonomy calculated over $\gamma_L \circ \gamma_0$ is identical to that calculated over $\gamma_t \circ \gamma_{x_R}$ and we have the identity
\begin{align}
    W_{\gamma_t} \ \cdot \ W_{\gamma_{x_R}} = W_{\gamma_L} \ \cdot \ W_{\gamma_0}.
    \label{eq:holomias1p1}
\end{align}

In terms of the physical fields, using (\ref{eq: wsol}) with $W_0 = 1$ for simplicity, the operator on the r.h.s. is given by
\begin{align}
    W_{\gamma_L} \ \cdot \ W_{\gamma_0} = e^{i2\beta \int_0 ^t j\big{|}_{x=L} dt'} e^{-i2\beta Q \vert_{t=0}} \nonumber
\end{align}
and similarly for the l.h.s.
\begin{align}
    W_{\gamma_t} \ \cdot \ W_{\gamma_{x_R}} = e^{-i2\beta Q\vert_{t>0}}e^{i2\beta \int_0 ^t j\vert_{x=x_R} dt'}, \nonumber
\end{align}
where $Q$ is the electric charge
\begin{equation}
Q = \int_{x_R}^{x_L} \rho \; dx.
\end{equation}

We can rewrite (\ref{eq:holomias1p1}) as
\begin{equation}
 W_{\gamma_t} = W_{\gamma_L} \ \cdot \ W_{\gamma_0} \ \cdot \  W_{\gamma_{x_R}}^{-1}
\end{equation}

and assuming that $j(t,x)\to 0$ in the limit where $x_R$ and $L$ go to infinity, the operators $W_{\gamma_L}$ and $W_{\gamma_{x_R}}$ become the identity and we remain with 
\begin{align}
    e^{-i2\beta Q(t)} = e^{-i2\beta Q(0)} \Rightarrow e^{-i2\beta \Delta Q} =1
\end{align}
where $\Delta Q \equiv Q \vert_{t>0} - Q \vert_{t=0}$. Since $\beta$ is arbitrary this identity implies 
\begin{align}
    \Delta Q =0
\end{align}
giving us the conservation of electric charge in time. These charges are defined as the eigenvalues of the holonomy restricted to space
\begin{equation}
    W_{\gamma_t} = e^{-i2\beta Q\vert_{t}}.
\end{equation}

This conservation law is a consequence of the path invariance of the holonomy operator or equivalently, of the flatness of the connection in loop space.


\begin{section}{The integral equations of electromagnetism in 2+1 dimensions}


In $3$-dimensional space-time the differential equations of electrodynamics are given by\cite{boito}
\begin{eqnarray}
    \partial_\mu F^{\mu \nu} &=& \frac{2\pi}{c}J^{\nu}\label{eq: maxwell3d1}\\ 
    \partial_\mu \widetilde{F}^{\mu}&=& 0\label{eq: maxwell3d2}, \qquad\mu,\nu = 0,1,2
\end{eqnarray}
where $J^{\mu} = (c\rho , j^1 , j^2)$ is the electric $3$-current density, $F_{\mu\nu} = \partial_\mu a_\nu - \partial_\nu a_\mu$ is the electromagnetic field whose components are the electric vector field $\mathbf{E}$ and the magnetic scalar field $B$ given as $F_{0i} = E_i$ and $F_{ij}=-\epsilon_{ij}B$, and $\widetilde{F}^{\mu} =\frac{1}{2} \epsilon^{\mu\nu\lambda}F_{\nu \lambda}$ is the Hodge dual of the electromagnetic field\footnote{We use $\epsilon^{012} = 1$ and $\epsilon_{ij} \equiv \epsilon^{0ij}$.}.

The integral equations of electromagnetism will be represented in the loop space $\mathcal{L}^{(1)}(\mathcal{M})$ which is defined by the maps $\mathcal{L}^{(1)}(\mathcal{M}) = \left\{\Gamma: S^1 \to \mathcal{M} \vert \Gamma(0) = x_R\right\}$. We consider a reference point $x_R$ in space-time $\mathcal{M}$ and a family of loops (closed curves) based at this point. Then, each of these loop will correspond to a point in loop space, which is the image of the map defined above. So, a collection of homotopic loops scanning a $2$-dimensional surface will define, in the loop space, a path $\Gamma$.

\begin{figure}[h]
    \centering
    \includegraphics[scale=0.4]{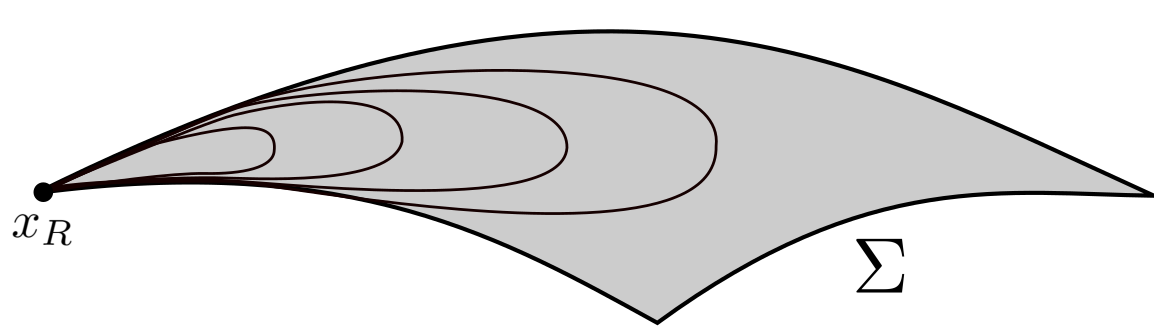}
    \caption{The scanning of the surface $\Sigma$ with loops based at $x_R$. Each loop in this scanning corresponds to a point in $\mathcal{L}^{(1)} (\mathcal{M})$.}
    \label{fig11}
\end{figure}

We define a $1$-form connection in the loop space as
\begin{align}
    \mathcal{A} =\oint_{\gamma} G_{\mu\nu}\frac{\partial x^\mu}{\partial\sigma}\delta x^\nu d\sigma
    \label{eq:conexao2p1}
\end{align}
where $G_{\mu\nu}$ is an anti-symmetric tensor which we integrate over each loop based at $x_R$ parameterized by $\sigma \in [0,2\pi]$ and labelled by $\tau \in [0,2\pi]$. So, this connection is defined, in space-time, on each loop and thus, it takes values at the points of the loop space $\mathcal{L}^{(1)}(\mathcal{M})$.

Taking $G_{\mu\nu} = \partial_\mu C_\nu -\partial_\nu C_\mu$, the components of the exact $2$-form $G=dC$, this connection becomes flat, i.e, its curvature vanishes: $\delta \mathcal{A} = 0$. So, in order to find an integral representation of the zero-curvature equation as the integral equations of the electromagnetism in loop space we write the components of the $1$-form $C=C_\mu dx^\mu$ in terms of the physical fields as
\begin{align}
    C_\mu = i\left(eA_\mu +\beta\widetilde{F}_\mu\right)
    \label{eq:2p1C}
\end{align}
where $\beta$ is an arbitrary constant and $e$ the elementary electric charge.

Using the dynamical equations (\ref{eq: maxwell3d1}) and (\ref{eq: maxwell3d2}) the components of $G$ can be written as
\begin{align}
    G_{\mu\nu} =  ieF_{\mu\nu}  + \frac{i2\pi \beta}{c} \epsilon_{\mu\nu\lambda}J^{\lambda}  \label{eq:2p1G}
\end{align}
and the connection in loop space reads
\begin{align}
    \mathcal{A} = ie\oint_\gamma F_{\mu\nu} \frac{\partial x^{\mu}}{\partial \sigma} \delta x^{\nu} d\sigma +\frac{i2\pi \beta}{c}\oint_\gamma \epsilon_{\mu\nu\lambda}J^{\lambda} \frac{\partial x^\mu}{\partial \sigma} \delta x^{\nu}d\sigma.
\end{align}

As in the previous case, the representation of the integral equations in loop space will be defined by an equation like (\ref{eq: loopspace_1d}),
\begin{equation}
\Delta \varphi= \int_{\Gamma}\mathcal{A}(\tau)d\tau
\end{equation}
where $\Gamma$, the path in loop space, stands now for the scanning of the $2$-dimensional surface $\Sigma$ in space-time with homotopically equivalent loops, based at $x_R$ which is located at the border $\partial \Sigma$ and 
\begin{equation}
\mathcal{A}(\tau) \equiv \int_0^{2\pi} G_{\mu\nu}\frac{\partial x^{\mu}}{\partial \sigma}\frac{\partial x^{\nu}}{\partial \tau}d\sigma.
\end{equation}

This integral equation in loop space is in fact a representation of the Stokes theorem in space-time for the 1-form $C$, which is in this case the potential $\varphi$:
\begin{align}
\label{eq: stokes1form}
    \oint_{\partial \Sigma}C_\mu\frac{dx^\mu}{d\sigma}d\sigma = - \int_\Sigma G_{\mu\nu} \frac{\partial x^\mu}{\partial \sigma} \frac{\partial x^\nu}{\partial \tau}d\sigma d\tau,
\end{align}
where $\Sigma$ is a 2-dimensional surface in spacetime and $\tau\in [0,2\pi]$ parameterizes the loops scanning this surface such that the loop with $\tau=0$ is the infinitesimal one (or point-loop) around the reference point $x_R$ and the loop with $\tau=2\pi$ is that which defines the border $\partial \Sigma$ of the surface $\Sigma$.

The integral equations of electromagnetism are a consequence of the Stokes theorem and the differential equations of motion, so, writing the fields in (\ref{eq: stokes1form}) as defined by (\ref{eq:2p1C}) and (\ref{eq:2p1G}), given the arbitrariness of $\beta$ we have the set of equations
\begin{align}
    &\oint A_\mu \frac{dx^\mu}{d\sigma}d\sigma = -\int_\Sigma F_{\mu\nu}\frac{\partial x^{\mu}}{\partial \sigma}\frac{\partial x^{\nu}}{\partial \tau}d\sigma d\tau
\label{2p1integral1}\\
    &\oint_{\partial\Sigma}\widetilde{F}_\mu \frac{dx^\mu}{d\sigma}d\sigma = -\frac{2\pi}{c}\int_\Sigma \epsilon_{\mu \nu \lambda}J^{\lambda}\frac{\partial x^\mu}{\partial \sigma}\frac{\partial x^\nu}{\partial \tau}d\sigma d\tau.
    \label{2p1integral2}
\end{align}

Let us show that this set of Lorentz scalar integral equations imply those usually presented, namely, the Gauss law, the Maxwell-Ampère law and the Faraday law.

The Gauss law is obtained from (\ref{2p1integral2}) when we take the surface $\Sigma$ to be completely spatial. For the l.h.s of this equation we get
\begin{align}
    \oint_{\partial\Sigma}\widetilde{F}_\mu \frac{dx^\mu}{d\sigma}d\sigma = \oint_{\partial\Sigma}\widetilde{F}_i \frac{dx^i}{d\sigma}d\sigma = \oint_{\partial\Sigma} \epsilon_{ij }E_j \frac{dx^i}{d\sigma}d\sigma \nonumber
\end{align}
and defining the normal vector to the curve $\partial \Sigma$ as $\hat{\mathbf{n}}_i \ dr = \epsilon_{ij}dx^j$ we can write the above result as
\begin{align}
    \oint_{\partial\Sigma}\widetilde{F}_\mu \frac{dx^\mu}{d\sigma}d\sigma = - \oint_{\partial\Sigma}\mathbf{E}\cdot \hat{\mathbf{n}} \ dr.
\end{align}

For the r.h.s. of (\ref{2p1integral2}) we have
\begin{align}
    & -\frac{2\pi}{c}\int_\Sigma \epsilon_{\mu \nu \lambda}J^{\lambda}\frac{\partial x^\mu}{\partial \sigma}\frac{\partial x^\nu}{\partial \tau}d\sigma d\tau=-\frac{2\pi}{c}\int_\Sigma \epsilon_{ij0}J^{0}\frac{\partial x^i}{\partial \sigma}\frac{\partial x^j}{\partial \tau}d\sigma d\tau \nonumber \\
   &=-2\pi \int_\sigma \rho \epsilon_{ij}\frac{\partial x^i}{\partial \sigma}\frac{\partial x^j}{\partial \tau}d\sigma d\tau = -2\pi \int_\sigma \rho \ dS
\end{align}
where $dS = \epsilon_{ij}\frac{\partial x^i}{\partial \sigma}\frac{\partial x^j}{\partial \tau}d\sigma d\tau$ is the area element.

Then the integral equation (\ref{2p1integral2}) becomes the usual Gauss' law for the electric field:
\begin{align}
    \oint_{\partial\Sigma}\mathbf{E}\cdot \hat{\mathbf{n}} \ dr=2\pi \int_\Sigma \rho \ dS.
\end{align}

Now, consider the 2-dimensional surface $\Sigma$ with a component in the $x^0$ direction as depicted in figure \ref{fig2}. In this case, the l.h.s. of (\ref{2p1integral2}) reads
\begin{align}
    \oint_{\partial\Sigma}\widetilde{F}_\mu \frac{dx^\mu}{d\sigma}d\sigma &= \int_{\gamma_0} \widetilde{F}_i \big|_{t=0}dx^i + \int_{\gamma_b} \widetilde{F}_0 \big|_{b}dx^0 - \int_{\gamma_t} \widetilde{F}_i \big|_{t>0}dx^i - \int_{\gamma_a} \widetilde{F}_0 \big|_{a}dx^0 \nonumber \\
    & = \int_{\gamma_0} \epsilon_{ij}E_j  \big|_{t=0}dx^i - \int_{\gamma_b} cB \big|_{b}dt - \int_{\gamma_t} \epsilon_{ij}E_j \big|_{t>0}dx^i + \int_{\gamma_a} cB \vert_{a}dt \nonumber \\
    &=\int_{\gamma_t} \mathbf{E}\vert_{t>0} \cdot \hat{\mathbf{n}} \  dr - \int_{\gamma_0} \mathbf{E}\big|_{t=0} \cdot \hat{\mathbf{n}} \  dr - c \left( \int_{\gamma_b} B \vert_{b}dt  -\int_{\gamma_a} B \vert_{a}dt \right)
\end{align}
and the r.h.s. is
\begin{align}
     -\frac{2\pi}{c}\int_\Sigma \epsilon_{\mu \nu \lambda}J^{\lambda}\frac{\partial x^\mu}{\partial \sigma}\frac{\partial x^\nu}{\partial \tau}d\sigma d\tau&=  -\frac{2\pi}{c}\int_{0}^{2\pi}\int_0^{2\pi} \epsilon_{i0j}J^{j}\frac{\partial x^i}{\partial \sigma}\frac{\partial x^0}{\partial \tau}d\sigma d\tau
     = -2\pi \int_0 ^t\int_{a}^b\mathbf{j} \cdot \hat{\mathbf{n}}\ dr dt'.
\end{align}

\begin{figure}[h]
    \centering
    \includegraphics[scale=0.6]{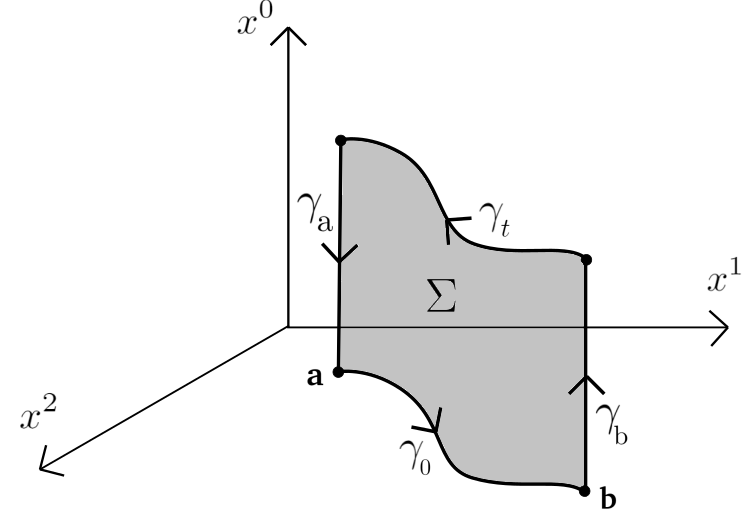}
    \caption{In $2+1$ dimensions we consider a 2-dimensional surface in space-time in order to obtain the Amp\`ere-Maxwell law from the Lorentz scalar integral equation.}
    \label{fig2}
\end{figure}

If we consider an infinitesimal time lapse in the results above then the equation (\ref{2p1integral2}) becomes
\begin{align}
    \int_{\gamma_t}\mathbf{E}\big|_{t>0} \cdot \hat{\mathbf{n}} \ dr - \int_{\gamma_0}\mathbf{E}\big|_{t=0} \cdot \hat{\mathbf{n}} \ dr - c\Delta t \left(  B \big|_{b}  - B \big|_{a} \right) = -2\pi \Delta t \int_a ^{b} \mathbf{j}\cdot \hat{\mathbf{n}} \ dr 
\end{align}
and in the limit where $\Delta t \to 0 $ we have finally the usual Amp\`ere-Maxwell law

\begin{align}
    B \big|_{b}  - B \big|_{a} = \frac{1}{c}\frac{d}{dt} \int_a ^b \mathbf{E} \cdot \hat{\mathbf{n}} \ dr +2\pi \int_a ^b \mathbf{j} \cdot \hat{\mathbf{n}} \ dr.
\end{align}

The Faraday law is in fact a mathematical identity and this is clear from the integral version given by the Lorentz scalar expression (\ref{2p1integral1}). In order to obtain the usual formula of this law we consider a 2-dimensional spatiotemporal surface defined by folding the previously used open surface such that $\gamma_a \sim \gamma_b ^{-1}$. 

The l.h.s. of (\ref{2p1integral1}) reads
\begin{align*}
    \oint A_\mu dx^\mu = \oint_{\gamma_0} A_i\big|_{t=0} dx^i- \oint_{\gamma_t} A_i\big|_{t>0} dx^i
\end{align*}
and using (\ref{2p1integral1}) again for each of the terms above at constant time we get
\begin{align}
     \oint A_\mu dx^\mu = \int_{S_0} B\big|_{t=0} dS - \int_{S_t} B\big|_{t>0} dS 
\end{align}
where $S_0$ and $S_t$ are the areas enclosed by $\gamma_0$ and $\gamma_t$, respectively.

Now, for the r.h.s of (\ref{2p1integral1}) we have
\begin{align}
    -\int_\Sigma F_{\mu\nu}\frac{\partial x^{\mu}}{\partial \sigma}\frac{\partial x^{\nu}}{\partial \tau}d\sigma d\tau &= - \int_0^{2\pi} \int_0^{2\pi} F_{i0}\frac{\partial x^{i}}{\partial \sigma}\frac{\partial x^{0}}{\partial \tau}d\sigma d\tau \nonumber \\
    &=\oint_{\partial S_0} \int_0 ^{t} c \ E_i dt' dx^i.
\end{align}

Considering an infinitesimal time lapse in these expressions, equation (\ref{2p1integral1}) becomes
\begin{align}
    \int_{S_0} B\big|_{t=0} dS - \int_{S_t} B\big|_{t>0} dS = -c\Delta t \oint_{\partial S_0} \mathbf{E} \cdot d\mathbf{r} \nonumber
\end{align}
which, in the limit of $\Delta t \to 0$ becomes
\begin{align}
    \oint_{\partial S_0} \mathbf{E} \cdot d\mathbf{r} = - \frac{1}{c}\frac{d}{dt}\int_{S_0} B dS.
\end{align}

\begin{figure}[h]
    \centering
    \includegraphics[scale=0.5]{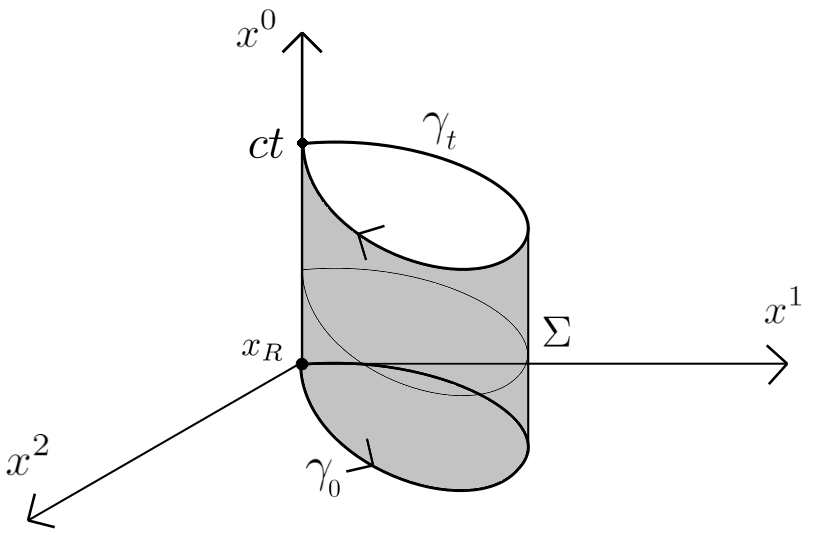}
    \caption{The 2-dimensional sheet in space-time from figure \ref{fig2} is folded to give a cylinder on which the Lorentz scalar integral equation is evaluated, resulting in the Faraday law. }
    \label{fig3}
\end{figure}

\subsection{The conservation of the electric charge in 3 dimensions}

The loop space connection is flat and this result is a direct consequence of the fact that $G$ is exact so $dG =0$ and
\begin{align}
    \delta \mathcal{A} = \oint \left(\partial_\lambda G_{\mu\nu} +\partial_\mu G_{\nu\lambda}+\partial_\nu G_{\lambda\mu}\right) \frac{\partial x^{\lambda}}{\partial \sigma} \delta x^{\mu} \land \delta x^{\nu}d\sigma =0 .
    \label{}
\end{align}

In terms of the physical fields, we find that
\begin{align}
 \delta \mathcal{A} =  \frac{i2\pi\beta}{c} \int \partial_\mu J^{\mu}\;d^3x =0 
\end{align}
and consequently the electric charge is conserved. 

These conserved charges can be obtained from the generalization of the holonomy operator defined by the $2$-holonomy $V$: a parallel transport operator in loop space $\mathcal{L}^{(1)}(\mathcal{M})$ obeying the equation\footnote{The reasoning behind the definition of this equation is explained in the appendix \ref{app}}
\begin{align}
    \frac{dV}{d\tau} - \left(\int_0 ^{2\pi}G_{\mu\nu} \frac{\partial x^{\mu}}{\partial \sigma}\frac{\partial x^{\nu}}{\partial \tau}d\sigma \right)V = 0
    \label{eq:2holonomia}
\end{align}
whose solution can be formally written as
\begin{align}
    V_\Sigma = e^{\int_\Sigma G_{\mu\nu} \frac{\partial x^\mu}{\partial \sigma}\frac{\partial x^\nu}{\partial \tau}d\sigma d\tau} \ V_{0}
\end{align}
where $\Sigma$ is a $2$-dimensional surface and $V_0$ is obtained from the initial conditions.

\begin{figure}[h]
    \centering
    \includegraphics[scale=0.4]{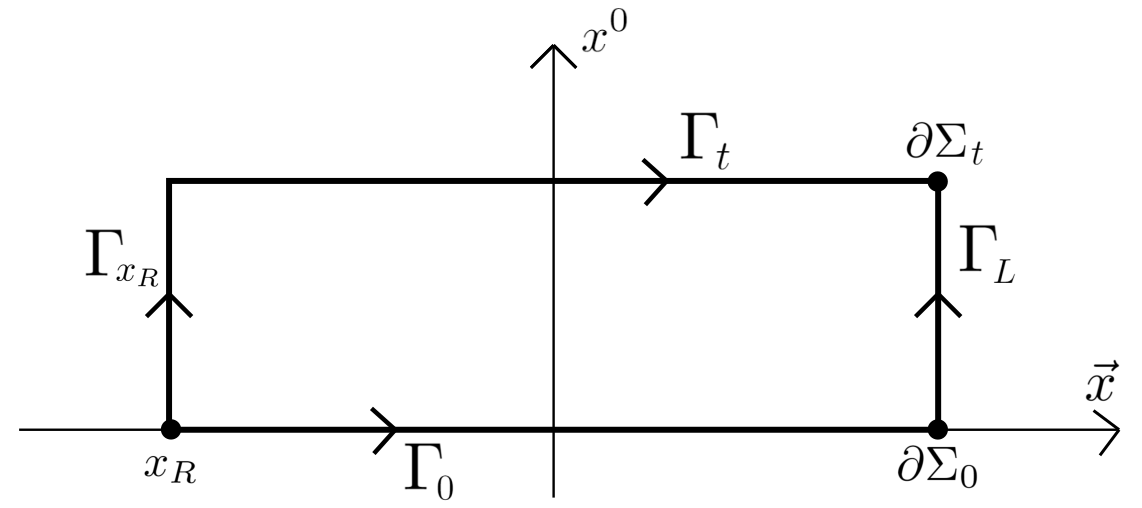}
    \caption{The flatness of the connection $\mathcal{A}$ implies that the parallel transport along any surface between the point-loop and $\partial \Sigma_t$ will give the same result. This can be defined in loop space in terms of the path-independence of the $2$-holonomy $V$.}
    \label{fig4}
\end{figure}

We consider the path $\Gamma_L \circ \Gamma_0$ in loop space, as in figure \ref{fig4}: the spatial surface $\Sigma_0 \sim \Gamma_0$ at constant time $t=0$ is scanned with loops starting at the point-loop at $x_R$, until the loop which defines the border $\partial \Sigma_0$ and then, moving from this last loop forward in time up to the loop $\partial \Sigma_t$.  The $2$-holonomy along this path (over this surface) is given by (considering, for simplicity, $V_0 = 1$)
\begin{align}
    V_{\Gamma_L} \ \cdot\  V_{\Gamma_0} = e^{\int_{\partial \Sigma_0 \times \mathbf{R}} G_{i0}\frac{\partial x^i}{\partial \sigma}\frac{\partial x^0}{\partial \tau}d\sigma d\tau}  e^{\int_{ \Sigma_0 } G_{ij}\frac{\partial x^i}{\partial \sigma}\frac{\partial x^j}{\partial \tau}d\sigma d\tau}.
\end{align}

From (\ref{eq:2p1G}) we have that
\begin{eqnarray*}
    \int_{\partial \Sigma_0 \times \mathbf{R}} G_{i0}\frac{\partial x^i}{\partial \sigma}\frac{\partial x^0}{\partial \tau}d\sigma d\tau &=& \int_{\partial \Sigma_0 \times \mathbf{R}}\left( ieF_{i0} + \frac{i2\pi \beta}{c}\epsilon_{i0j} J^j\right) \frac{\partial x^i}{\partial \sigma}\frac{\partial x^0}{\partial \tau}d\sigma d\tau \\
    &=&-iec\int \oint \mathbf{E}\cdot d\mathbf{r}dt + i 2\pi \beta \int \oint \mathbf{j}\cdot \hat{\mathbf{n}} drdt.
\end{eqnarray*}

Assuming that $\Vert\mathbf{E}\Vert$ falls off quickly enough and that the electric current is localized, the quantity above should vanish in the limit where the radius of $\partial \Sigma_0$ goes to infinity. What remains is then
\begin{align}
    V_{\Gamma_L} \ \cdot\  V_{\Gamma_0} = e^{ie\int_{\Sigma_0 } F_{ij}\frac{\partial x^i}{\partial \sigma}\frac{\partial x^j}{\partial \tau}d\sigma d\tau} e^{\frac{i2\pi  \beta}{c}\int_{ \Sigma_0 } \epsilon_{ij}J^{0}\frac{\partial x^i}{\partial \sigma}\frac{\partial x^j}{\partial \tau}d\sigma d\tau}
    = e^{-ie\int_{\Sigma_0}B\big|_{t=0}dS} e^{i2\pi \beta Q\big|_{t=0}}.
\end{align}

In the construction of the $2$-holonomy over the path $\Gamma_t \circ \Gamma_{x_R}$ we notice that $V_{\Gamma_{x_R}}$ becomes trivial as the radius of the point-loop around $x_R$ becomes zero and we get
\begin{align}
     V_{\Gamma_t} \ \cdot\  V_{\Gamma_{x_R}} = e^{-ie\int_{\Sigma_0}B\big|_{t>0}dS} \ \cdot\ e^{i2\pi \beta Q\big|_{t>0}}.
\end{align}

The path independence of the $2$-holonomy is a direct consequence of the flatness of the connection in loop space and it implies that, once $\Gamma_t \circ \Gamma_{x_R}$ can be obtained from continous deformations of $\Gamma_L \circ \Gamma_0$, we have the relation
\begin{align}
     V_{\Gamma_t} \ \cdot\  V_{\Gamma_{x_R}} =  V_{\Gamma_L} \ \cdot\  V_{\Gamma_0} \qquad \Rightarrow \qquad      V_{\Gamma_t} =  V_{\Gamma_L} \ \cdot\  V_{\Gamma_0} \ \cdot \  V_{\Gamma_{x_R}}^{-1}.
\end{align}

Given that the operators $V_{\Gamma_L}$ and $V_{\Gamma_{x_R}}$, as discussed above, become unity, the path independence of the $2$-holonomy gives us that the electric charge $Q$, defined by the eigenvalues of
\begin{equation}
    V_{\Gamma_t} = e^{i 2\pi \beta Q\vert_t}
\end{equation}
and the magnetic flux over the entire space, defined by the eigenvalues of
\begin{equation}
    V_{\Gamma_t} = e^{-ie\int_{\Sigma_0}B\vert_t\;dS}
\end{equation}
are conserved in time.
\end{section}

 \section{The integral equations of electromagnetism in $3+1$ dimensions}

In $3+1$ dimensions, the electric and magnetic vector fields $\mathbf{E}$ and $\mathbf{B}$ can be written as the components of the electromagnetic field strength $F_{\mu\nu}$, $\mu,\nu = 0,\dots,3$ as $E^i = F_{0i}=-F_{i0}$ and $B^k = -\frac{1}{2}\epsilon_{kij}F_{ij}$.

Maxwell equations are given by 
\begin{eqnarray}
\partial_\mu F^{\mu\nu} &=& \frac{4\pi}{c}J^\nu \label{eq: maxeq}\\
\partial_\mu \widetilde{F}^{\mu\nu}&=&0 \label{eq: bianchi}
\end{eqnarray}
where $J^\mu = (c\rho, j^i)$ is the Lorentz covariant current and $ \widetilde{F}^{\mu\nu} = \frac{1}{2}\epsilon^{\mu\nu\lambda\gamma}F_{\lambda\gamma}$ is the Hodge dual of the electromagnetic field.

The construction of the integral representation of the Maxwell equation is done in the loop space $\mathcal{L}^{(2)}(\mathcal{M})= \left\{\Gamma: S^2 \to \mathcal{M}\vert \Gamma(0) = x_R\right\}$. This mapping will relate closed $2$-dimensional surfaces in space-time, which are all based at the reference point $x_R$, with points in the loop space. 

\begin{figure}[ht!]
    \centering
    \includegraphics[scale=0.5]{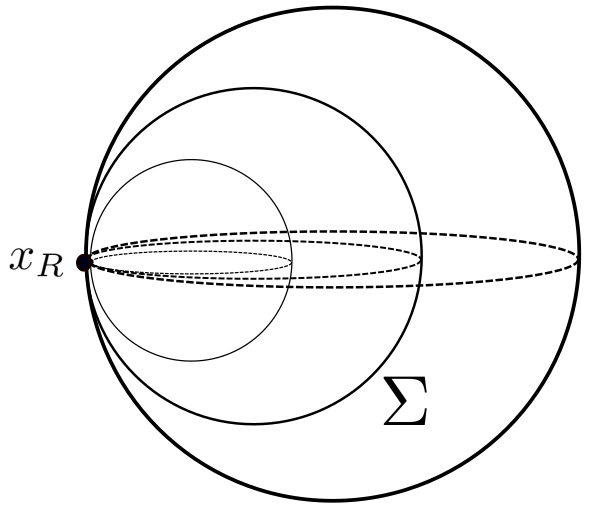}
    \caption{The scanning of space-time is doing by considering a family of spheres based at $x_R$ which can be continuously deformed into each other.}
    \label{fig12}
\end{figure}

We define the $1$-form connection in $\mathcal{L}^{(2)}(\mathcal{M})$ by
\begin{equation}
    \mathcal{A} = \oint_{\Sigma}H_{\mu\nu\lambda}\frac{\partial x^\mu}{\partial \sigma}\frac{\partial x^\nu}{\partial \tau}\delta x^\lambda d\sigma d\tau
\end{equation}
where $H_{\mu\nu\lambda}$ is a $3$-form in space-time which is integrated over the $2$-dimensional closed surface $\Sigma$, parameterized by $\sigma \in [0,2\pi]$ and $\tau \in [0,2\pi]$. By scanning a $3$-dimensional volume in space-time with these closed surfaces as in figure \ref{fig12} we define a path in the loop space.

Let us take $H = dB$, i.e. we consider it to be an exact $3$-form. This is a sufficient condition for $\mathcal{A}$ to be flat, i.e., $\delta \mathcal{A} = 0$. We look for an integral representation in loop space of this zero-curvature equation such that it will be equivalent to Maxwell integral equations of electromagnetism. Such an integral equation in loop space is given by
\begin{equation}
\Delta \varphi = \int_\Gamma \mathcal{A}(\zeta)d\zeta
\end{equation}
with
\begin{equation}
 \mathcal{A}(\zeta) = \oint_{\Sigma}H_{\mu\nu\lambda}\frac{\partial x^\mu}{\partial \sigma}\frac{\partial x^\nu}{\partial \tau}\frac{\partial x^\lambda}{\partial \zeta} d\sigma d\tau
\end{equation}
and the potential $\varphi$ will be given by 
\begin{equation}
\varphi = \int_\Sigma B_{\mu\nu}\frac{\partial x^\mu}{\partial \sigma}\frac{\partial x^\nu}{\partial \tau}d\sigma d\tau.
\end{equation}

As before, the integral equation in loop space is a representation of the Stokes theorem for the $2$-form $B = \frac{1}{2}B_{\mu\nu}dx^\mu\wedge dx^\nu$:

\begin{equation}
\label{eq: stokes}
\oint_{\partial \Omega}B_{\mu\nu}\frac{\partial x^\mu}{\partial \sigma}\frac{\partial x^\nu}{\partial \tau}d\sigma d\tau = \int_\Omega \left(\partial_\lambda B_{\mu\nu}+\partial_\mu B_{\nu\lambda}+\partial_\lambda B_{\mu\nu}\right)\frac{\partial x^\mu}{\partial \sigma}\frac{\partial x^\nu}{\partial \tau}\frac{\partial x^\lambda}{\partial \zeta} d\sigma d\tau d\zeta.
\end{equation}

We now define the components $B_{\mu\nu}$ in terms of the electromagnetic field strength and its Hodge dual
\begin{equation}
\label{eq: b}
B_{\mu\nu} = i(\alpha F_{\mu\nu}+\beta \widetilde{F}_{\mu\nu})
\end{equation}
where $\alpha$ and $\beta$ are arbitrary constants. Once the Maxwell equations (\ref{eq: maxeq}) are satisfied by these fields, the integrand in the r.h.s of (\ref{eq: stokes}) can be written as  
\begin{equation}
\partial_\lambda B_{\mu\nu}+\partial_\mu B_{\nu\lambda}+\partial_\lambda B_{\mu\nu} = i\alpha \left(\partial_\lambda F_{\mu\nu}+\partial_\mu F_{\nu\lambda}+\partial_\lambda F_{\mu\nu}\right) + i\beta \epsilon_{\mu \nu \lambda  \gamma}\partial_\rho F^{\rho \gamma} = \frac{4\pi\beta}{c} \epsilon_{\mu\nu\lambda\gamma}J^\gamma.
\end{equation}
and the mathematical relation (\ref{eq: stokes}) defines the Lorentz scalar integral equations of electrodynamics:
\begin{equation}
\oint_{\partial \Omega}\left(\alpha F_{\mu\nu}+\beta \widetilde{F}_{\mu\nu}\right)\frac{\partial x^\mu}{\partial \sigma}\frac{\partial x^\nu}{\partial \tau}d\sigma d\tau = \frac{4\pi\beta}{c}\int_\Omega J^\gamma\epsilon_{\mu\nu\lambda\gamma}\frac{\partial x^\mu}{\partial \sigma}\frac{\partial x^\nu}{\partial \tau}\frac{\partial x^\lambda}{\partial \zeta} d\sigma d\tau d\zeta.
\end{equation}

The arbitrariness of $\alpha$ and $\beta$ implies that the following two equations hold simultaneously

\begin{eqnarray}
\oint_{\partial \Omega}F_{\mu\nu}\frac{\partial x^\mu}{\partial \sigma}\frac{\partial x^\nu}{\partial \tau}d\sigma d\tau&=&0 \label{eq: int1}\\
\oint_{\partial \Omega}\widetilde{F}_{\mu\nu}\frac{\partial x^\mu}{\partial \sigma}\frac{\partial x^\nu}{\partial \tau}d\sigma d\tau&=&\frac{4\pi}{c}\int_\Omega J^\gamma\epsilon_{\mu\nu\lambda\gamma}\frac{\partial x^\mu}{\partial \sigma}\frac{\partial x^\nu}{\partial \tau}\frac{\partial x^\lambda}{\partial \zeta} d\sigma d\tau d\zeta.\label{eq: int2}
\end{eqnarray}

These are Lorentz scalar equations for the flux of the electromagnetic field strength and its Hodge dual through 2-dimensional surfaces in Minkowski space-time.

We now proceed to show that these equations imply the usual integral laws of electrodynamics when the 3-dimensional volumes in space-time are appropriately chosen. 

Let us start by considering equation (\ref{eq: int1}), which is the integral version of the Bianchi identity, when $\Omega$ is a 3-dimensional spatial volume at a given instant of time. The l.h.s of that equation becomes\footnote{Here the Hodge dual of $dx^i\wedge dx^j$ in $\mathbb{R}^3$ is given by $dS^k = \epsilon_{ijk}\frac{\partial x^i}{\partial \sigma}\frac{\partial x^j}{\partial \tau}d\sigma d\tau$.} 
\begin{eqnarray*}
\oint_{\partial \Omega}F_{\mu\nu}\frac{\partial x^\mu}{\partial \sigma}\frac{\partial x^\nu}{\partial \tau}d\sigma d\tau &=& \oint_{\partial \Omega}F_{ij}\frac{\partial x^i}{\partial \sigma}\frac{\partial x^j}{\partial \tau}d\sigma d\tau
 =-\oint_{\partial \Omega}B^k\epsilon_{ijk} \frac{\partial x^i}{\partial \sigma}\frac{\partial x^j}{\partial \tau}d\sigma d\tau\\
 &=& -\oint_{\partial \Omega}\mathbf{B}\cdot d\mathbf{S}
\end{eqnarray*}
and the r.h.s being equal to zero, we recover the Gauss law for the magnetic field.

Next we consider the same equation (\ref{eq: int1}) but now with $\Omega$ as a 3-dimensional volume in space-time which we shall take as the cylinder $\Omega = \mathbb{D}^2\times \mathbb{R}$. The l.h.s of equation (\ref{eq: int1}) is now decomposed in three parts, corresponding to the flux of the electromagnetic field strength across the three surfaces which form the border of the cylinder: $\partial \Omega = (\mathbb{D}^2_0)^{-1}\cup \mathbb{D}^2_t\cup (\mathbb{S}^1\times \mathbb{R})$. The time direction is taken at the axis of symmetry of this cylinder and we reverse the orientation of the bottom disk $\mathbb{D}^2_0$, at $t = 0$, so that we can consider it as a closed orientable surface.
\begin{figure}[h!]
  \centering
    \includegraphics[width=0.5\textwidth]{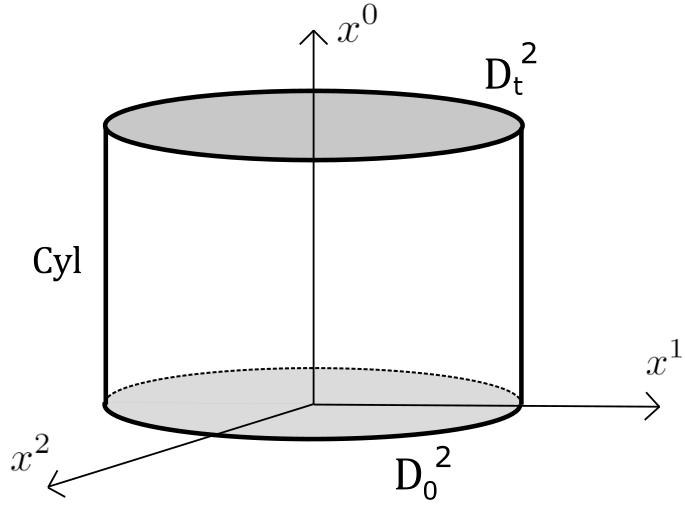}
    \caption{The cylinder defines a closed surface in space-time with its symmetry axis along the time dimension. The bottom ($\mathbb{D}^2_0$) and top ($\mathbb{D}^2_t$) disks are 2-dimensional spatial surfaces at different instants of time while the surface on the side of the cylinder (cyl $ \sim \mathbb{S}^1\times \mathbb{R}$) has two perpendicular directions in time and space.} 
    \label{fig: cilindro}
\end{figure}

Then we have, for the flux of the electromagnetic field on the cylinder:
\begin{eqnarray*}
\oint_{\partial \Omega}F_{\mu\nu}\frac{\partial x^\mu}{\partial \sigma}\frac{\partial x^\nu}{\partial \tau}d\sigma d\tau &=& -\oint_{\mathbb{D}^2_0}F_{ij}\frac{\partial x^i}{\partial \sigma}\frac{\partial x^j}{\partial \tau}d\sigma d\tau +\oint_{\mathbb{D}^2_t}F_{ij}\frac{\partial x^i}{\partial \sigma}\frac{\partial x^j}{\partial \tau}d\sigma d\tau + \oint_{\mathbb{S}^1\times \mathbb{R}}F_{\mu\nu}\frac{\partial x^\mu}{\partial \sigma}\frac{\partial x^\nu}{\partial \tau}d\sigma d\tau.
\end{eqnarray*}

The two first integrals will give the flux of the magnetic field through each of the disks, at different instants of time. For the third integral we take $\sigma$ to parameterize the surface of the cylinder at constant $x^0=ct$ coordinate which in turn is parameterized with $\tau$. Then
\[
\oint_{\mathbb{S}^1\times \mathbb{R}}F_{\mu\nu}\frac{\partial x^\mu}{\partial \sigma}\frac{\partial x^\nu}{\partial \tau}d\sigma d\tau = -c \int_0^{t} \int_0^{2\pi} E^i\frac{\partial x^i}{\partial \sigma}d\sigma dt'
\]
and if we assume an infinitesimal time lapse, the l.h.s of equation (\ref{eq: int1}) reads
\[
\oint_{\partial \Omega}F_{\mu\nu}\frac{\partial x^\mu}{\partial \sigma}\frac{\partial x^\nu}{\partial \tau}d\sigma d\tau  = -\Delta \Phi(\mathbf{B},\mathbb{D}^2) - c\Delta t \oint_{\partial \mathbb{D}^2}\mathbf{E}\cdot d\mathbf{x}
\]
where $\Delta \Phi(\mathbf{B},\mathbb{D}^2)$ stands for the change in the flux of magnetic field from $\mathbb{D}^2_0$ to $\mathbb{D}^2_t$. Finally, the integral Bianchi identity becomes
\[
\frac{1}{c}\frac{\Delta \Phi(\mathbf{B},\mathbb{D}^2)}{\Delta t} +  \oint_{\partial \mathbb{D}^2}\mathbf{E}\cdot d\mathbf{x} = 0
\]
which, in the limit $\Delta t \to 0$ gives the Faraday law:
\begin{equation}
\frac{1}{c}\frac{d \Phi(\mathbf{B},\mathbb{D}^2)}{d t} +  \oint_{\partial \mathbb{D}^2}\mathbf{E}\cdot d\mathbf{x} = 0.
\end{equation}

Now we consider equation (\ref{eq: int2}), which corresponds to the Maxwell equations with the matter sources. Taking $\Omega$ to be a 3-dimensional spatial volume at a given time, its l.h.s becomes
\begin{eqnarray*}
\oint_{\partial \Omega}\widetilde{F}_{\mu\nu}\frac{\partial x^\mu}{\partial \sigma}\frac{\partial x^\nu}{\partial \tau}d\sigma d\tau = \oint_{\partial \Omega}\widetilde{F}_{ij}\frac{\partial x^i}{\partial \sigma}\frac{\partial x^j}{\partial \tau}d\sigma d\tau = -\oint_{\partial \Omega}E^k\epsilon_{kij}\frac{\partial x^i}{\partial \sigma}\frac{\partial x^j}{\partial \tau}d\sigma d\tau = -\oint_{\partial \Omega}\mathbf{E}\cdot d\mathbf{S},
\end{eqnarray*}
i.e., the flux of the electric field across the border of that spatial volume. The r.h.s of (\ref{eq: int2}), when evaluated in that spatial volume, reads
\[
\frac{4\pi}{c}\int_\Omega J^\gamma\epsilon_{\mu\nu\lambda\gamma}\frac{\partial x^\mu}{\partial \sigma}\frac{\partial x^\nu}{\partial \tau}\frac{\partial x^\lambda}{\partial \zeta} d\sigma d\tau d\zeta = \frac{4\pi}{c}\int_\Omega J^0\epsilon_{ijk0}\frac{\partial x^i}{\partial \sigma}\frac{\partial x^j}{\partial \tau}\frac{\partial x^k}{\partial \zeta} d\sigma d\tau d\zeta = -4\pi\int_\Omega \rho d^3\mathbf{x}
\]
and equation (\ref{eq: int2}) gives the Gauss law for the electric field.

Finally, considering $\Omega = \mathbb{D}^2\times \mathbb{R}$, that cylindrical volume in space and time, with the same parameterization as before, the l.h.s of (\ref{eq: int2}) becomes
\[
\oint_{\partial \Omega}\widetilde{F}_{\mu\nu}\frac{\partial x^\mu}{\partial \sigma}\frac{\partial x^\nu}{\partial \tau}d\sigma d\tau = \Phi(\mathbf{E},\mathbb{D}^2_0) - \Phi(\mathbf{E},\mathbb{D}^2_t) + c\int_0^t \int_0^{2\pi}B^i\frac{\partial x^i}{\partial \sigma} d\sigma dt',
\]
while its r.h.s reads
\[
\frac{4\pi}{c}\int_\Omega J^\gamma\epsilon_{\mu\nu\lambda\gamma}\frac{\partial x^\mu}{\partial \sigma}\frac{\partial x^\nu}{\partial \tau}\frac{\partial x^\lambda}{\partial \zeta} d\sigma d\tau d\zeta =\frac{4\pi}{c}\int_\Omega J^k\epsilon_{ij0k}\frac{\partial x^i}{\partial \sigma}\frac{\partial x^j}{\partial \tau}\frac{\partial x^0}{\partial \zeta} d\sigma d\tau d\zeta = 4\pi \int_0^t \int_{\mathbb{D}^2}j^k\epsilon_{kij}\frac{\partial x^i}{\partial \sigma}\frac{\partial x^j}{\partial \tau}d\sigma d\tau dt'. 
\]

For an infinitesimal time lapse, equation (\ref{eq: int2}) gives
\[
-\frac{1}{c}\frac{\Delta \Phi(\mathbf{E},\mathbb{D}^2)}{\Delta t} +  \oint_{\partial \mathbb{D}^2}\mathbf{B}\cdot d\mathbf{x} = \frac{4\pi}{c}  \Phi(\mathbf{j},\mathbb{D}^2),
\]
which in the limit $\Delta t \to 0$ defines the Amp\`ere-Maxwell law of induction:
\begin{equation}
-\frac{1}{c}\frac{d \Phi(\mathbf{E},\mathbb{D}^2)}{d t} +  \oint_{\partial \mathbb{D}^2}\mathbf{B}\cdot d\mathbf{x} = \frac{4\pi}{c}  \Phi(\mathbf{j},\mathbb{D}^2).
\end{equation}

\subsection{The conservation of the electric charge in 4 dimensions}

By construction, the loop space connection is flat, i.e, $\delta \mathcal{A} = 0$, since it is defined in terms of an exact field, $H = dB$ integrated over closed surfaces. This implies locally that the electric charge is conserved when we consider the definition of $B$ in terms of the physical fields and take into account the differential Maxwell equations:

\begin{equation}
\mathcal{A} = \frac{i4\pi\beta}{c}\oint_\Sigma J^\gamma \epsilon_{\mu\nu\lambda\gamma}\frac{\partial x^\mu}{\partial \sigma}\frac{\partial x^\nu}{\partial \tau} \delta x^\lambda d\sigma d\tau 
\end{equation}
and
\begin{equation}
\delta \int \mathcal{A} d\zeta  = -\frac{i 4\pi \beta}{c} \int \partial_\mu J^\mu d^4x = 0.
\end{equation}

In order to obtain the conserved charges we consider the generalization of the holonomy operator given by the $3$-holonomy $U$, satisfying the parallel transport equation
\begin{equation}
    \frac{dU}{d\zeta} - \int_{0}^{2\pi}\int_{0}^{2\pi} H_{\mu\nu\lambda}\frac{\partial x^\mu}{\partial \sigma}\frac{\partial x^\nu}{\partial \tau}\frac{ \partial x^\lambda}{\partial \zeta} d\sigma d\tau \; U= 0
\end{equation}
whose solution can be formally written as
\begin{equation}
    U_{\Omega} = e^{\int_{\Omega}H_{\mu\nu\lambda}\frac{\partial x^\mu}{\partial \sigma}\frac{\partial x^\nu}{\partial \tau}\frac{ \partial x^\lambda}{\partial \zeta} d\sigma d\tau d\zeta}\;U_0,
\end{equation}
where $U_0$ is defined by the initial conditions.

Clearly\footnote{See appendix \ref{app}.}, the flatness of the connection implies that the 3-holonomy operator is path-independent in loop space.

Now, we split space-time into space and time and construct a volume whose border changes in time from $t=0$. Since the 3-holonomy operator is path independent, let us use the following convenient path: a composition $\Gamma = \Gamma_L \circ \Gamma_0$ where $\Gamma_0$ starts at $\Sigma_0$ and goes to $\partial \Omega_0$, the border of the completely spatial volume $\Omega$ at constant time $t=0$ and $\Gamma_L$ a path which starts at $\partial \Omega_0$ and changes only in the time direction ending at $\partial \Omega_t$, the purely spatial volume at time $t$. 

\begin{figure}[h!]
  \centering
    \includegraphics[width=0.59\textwidth]{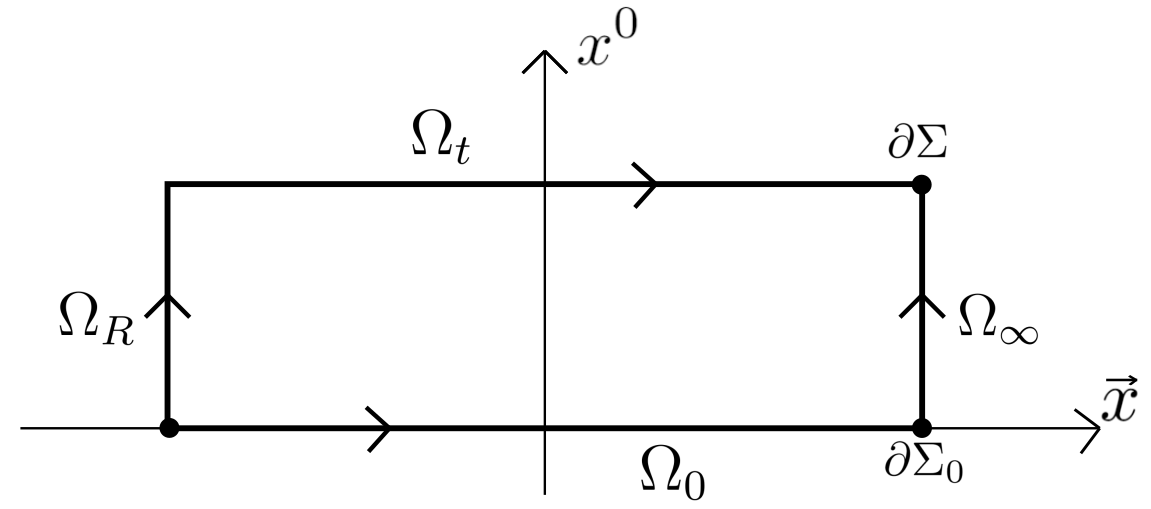}
    \caption{The path independence of the 3-holonomy defines the eingenvalues of this operator over a spatial volume as the conserved charges.} 
    \label{fig: pot_dens}
\end{figure}

The relevant quantity in the construction of the $3$-holonomy operator $U_\Gamma = U_{\Gamma_L}\cdot U_{\Gamma_0}$ is the integral of the connection in $\zeta$. Then we have
\begin{eqnarray}
\int_{\Gamma_0} \mathcal{A}(\zeta)d\zeta &=& \frac{i4\pi}{c} \beta \int_{\Omega_0}J^0 \epsilon_{ijk0}\frac{\partial x^i}{\partial \sigma}\frac{\partial x^j}{\partial \tau}\frac{\partial x^k}{\partial \zeta}d\sigma d\tau d\zeta = i4\pi \beta Q(0)\\
\int_{\Gamma_L} \mathcal{A}(\zeta)d\zeta &=& i4\pi \beta \int_0^t \int_{\partial \Omega}J^k \epsilon_{ij0k}\frac{\partial x^i}{\partial \sigma}\frac{\partial x^j}{\partial \tau}d\sigma d\tau  dt = i4\pi \beta \int_0^t \Phi(\mathbf{j},\partial \Omega)dt,
\end{eqnarray}
where $Q(0)$ stands for the electric charge at time $t=0$.

Now, assuming the charge distribution to be localized, then at $r\to \infty$ we have $\vert \mathbf{j}\vert \sim r^{-(2+\varepsilon)}$, $\varepsilon > 0$ and this means that $U_{\Gamma_L} = 1$ if the border $\partial \Omega_0$ is far enough from the charges. So,
we end up with (considering $U_0=1$ for simplicity)
\begin{equation}
U_\Gamma = e^{i4\pi \beta Q(0)}.
\end{equation}

Next we consider a similar path joining the same two points in loop space: $\Gamma' = \Gamma_t\circ \Gamma_{x_R}$ where now $\Gamma_t$ is the path joining the closed surface $\Sigma_0$ at time $t=0$ to the surface $\Sigma_t$ at a later time and $\Gamma_{x_R}$ is the purely spatial path joining the surface $\Sigma_t$ to the surface $\partial \Omega_t$, at constant time $t$. 

The operator $U_{\Gamma_{x_R}}$ will become the unity when the radius of the sphere at the reference point goes to zero and we remain with
\begin{equation}
U_{\Gamma'} = e^{i4\pi \beta Q(t)}.
\end{equation}


Now, the path-independence of $U$ dictates that $U_{\Gamma'} = U_\Gamma$, given that the two paths $\Gamma$ and $\Gamma'$ are homotopically equivalent in loop space. With the considerations above for the behaviour of the fields at spatial infinity, this relation gives
\begin{equation}
e^{i4\pi \beta \Delta Q}  = 1 \qquad \Delta Q \equiv Q(t)-Q(0)
\end{equation}
defining the conservation of the electric charge, which is given by the eigenvalues of the operator
\begin{equation}
\label{eq: u}
    U_{\Gamma_t} = e^{i4\pi \beta Q(t)}.
\end{equation}



\section{Conclusions}


We have shown that it is possible to formulate the integral equations of electrodynamics in $(D+1)$-dimensional space-time $\mathcal{M}$ as the integral version of the zero-curvature equation in the loop space $\mathcal{L}^{(D-1)}(\mathcal{M})$:
\begin{equation}
    \Delta \varphi = \int_{\Gamma}\mathcal{A}(s)ds
\end{equation}
where $\varphi$ defines the integral of a $(D-1)$-form over a $(D-1)$-dimensional hyper-surface in space-time and $\mathcal{A}$ is the corresponding flat connection given in terms of an exact $D$-form. In particular, for $D=1,2,3$ we have respectively
\begin{equation}
\varphi = f, \qquad 
\varphi = \int_\gamma C_\mu \frac{dx^\mu}{d\sigma}d\sigma, \qquad
\varphi = \int_{\Sigma}B_{\mu\nu}\frac{\partial x^\mu}{\partial \sigma}\frac{\partial x^\nu}{\partial \tau}d\sigma d\tau
\end{equation}
and 
\begin{equation}
\mathcal{A} = A_\mu \delta x^\mu,\qquad 
\mathcal{A} = \oint_\gamma G_{\mu\nu}\frac{\partial x^\mu}{\partial \sigma}\delta x^\nu d\sigma,\qquad
\mathcal{A} = \oint_\Sigma H_{\mu\nu\lambda}\frac{\partial x^\mu}{\partial \sigma}\frac{\partial x^\nu}{\partial \tau}\delta x^\lambda d\sigma d\tau
\end{equation}
with
\begin{equation}
    A = df,\qquad G = dC, \qquad H = dB.
\end{equation}

Generally speaking, the connection defined in $\mathcal{L}^{(D-1)}(\mathcal{M})$ is given in terms of the components of the $D$-form in space-time $\omega = \omega_{\mu_1\dots \mu_D}dx^{\mu_1}\wedge \dots \wedge dx^{\mu_D}$ as
\begin{equation}
\mathcal{A} = \oint_{\mathcal{K}}\omega_{\mu_1\dots \mu_D} \frac{\partial x^{\mu_1}}{\partial s_1}\dots \frac{\partial x^{\mu_{D-1}}}{\partial s_{D-1}}\delta x^{\mu_D}ds_1\dots ds_{D-1}. 
\end{equation}
This integral representation in loop space is equivalent to the differential equation
\begin{equation}
    \mathcal{A} = \delta \varphi.
\end{equation}
so that $\mathcal{A}$ is a pure gauge connection. The local conservation law for the electric charge becomes a consequence of the mathematical identity $\delta \mathcal{A} = \delta^2 \varphi = 0$ in loop space. The flatness of the connection in loop space implies that the generalized holonomy operator defined by
\begin{equation}
\frac{d\mathcal{W}}{ds} + (-1)^{(D-1)}\mathcal{A}(s)\;\mathcal{W} = 0
\end{equation}
whose solution can be formally written as
\begin{equation}
\mathcal{W} = e^{(-1)^D\int_\Gamma \mathcal{A}(s)ds}\mathcal{W}_\circ
\end{equation}
is path-independent.

A homotopic variation of $\Gamma$ implies a variation of $\mathcal{W}$ which depends on the curvature of $\mathcal{A}$ and so, $\delta\mathcal{W}= 0$. Therefore, the equations of electrodynamics have a symmetry under the homotopic transformations of the path in loop space and the conserved charges are obtained as the eigenvalues of the generalized holonomy, evaluated at constant time.

\vspace{2cm}
\textbf{Aknowledgments}

G. L. would like to thank L.A. Ferreira and H. Malavazzi for valuable discussions on the subject of the paper.

\appendix

\section{The construction of the generalized holonomies and their path-independence in loop space}
\label{app}

In what follows we present the reasoning behind the construction of the generalized holonomies which define the parallel transport operators in $\mathcal{L}^{(1)}(\mathcal{M})$ and $\mathcal{L}^{(2)}(\mathcal{M})$. This is a review but for the abelian case of what was first introduced in \cite{laf_1997} and later discussed in \cite{laf_2010} and \cite{luchini1}.

Consider the holonomy $W$ defined by the parallel transport equation
\begin{equation}
\label{app1}
\frac{dW}{d\sigma}+A_\mu \frac{dx^\mu}{d\sigma}W = 0
\end{equation}
along a curve $\gamma$. Suppose this curve to be closed and obtained by continuous deformations $x^\mu \to x^\mu + \delta x^\mu$ from a curve $\gamma_0$ sharing a common point $x_R$ with $\gamma$.

We label the set of homotopic closed curves with $\tau \in [0,2\pi]$ such that $\gamma$ has $\tau = 2\pi$ and $\gamma_0$, $\tau = 0$.

Let us assume that we know $W$ on $\gamma_0$. For instance, if $\gamma_0$ is the point-loop at $x_R$, W can be taken as $W_0$, an initial value for (\ref{app1}).

Now, if we want to get $W$ over $\gamma$, then instead of integrating (\ref{app1}) directly we can compute the change of $W$ which is calculated over $\gamma_0$ while this path is deformed into $\gamma$. 

For an infinitesimal deformation of the path we have $W \to W + \delta W$ and the variation $\delta W$ can be obtained from equation (\ref{app1}) as follows. We start by cosidering the variation of the equation as a whole:
\[
\delta \frac{dW}{d\sigma} + \delta \left(A_\mu \frac{dx^\mu}{d\sigma}\right)W + A_\mu \frac{dx^\mu}{d\sigma}\delta W = 0.
\]

Then we multiply this expression by $W^{-1}$ and rewrite the first term getting
\[
\frac{d}{d\sigma}\left(W^{-1}\delta W\right)+W^{-1}\frac{dW}{d\sigma}W^{-1}\delta W + \delta \left(A_\mu \frac{dx^\mu}{d\sigma}\right)+ W^{-1}A_\mu \frac{dx^\mu}{d\sigma}\delta W = 0
\]
and using equation (\ref{app1}) the second and fourth terms cancel each and what remains is
\[
\frac{d}{d\sigma}\left(W^{-1}\delta W\right) + \delta \left(A_\mu \frac{dx^\mu}{d\sigma}\right) = 0.
\]

This equation can be integrated in $\sigma \in [0,2\pi]$ giving
\[
\delta W = - \int_0^{2\pi} \delta \left(A_\mu \frac{dx^\mu}{d\sigma}\right)d\sigma
\]
and finally, calculating the variation of the term in the r.h.s:
\[
\delta W = - \int_0^{2\pi} \left(\partial_\nu A_\mu \frac{dx^\mu}{d\sigma}\delta x^\nu - \partial_\nu A_\mu \frac{dx^\nu}{d\sigma}\delta x^\mu\right)d\sigma - A_\mu \delta x^\mu\Big\vert_0^{2\pi}.
\]

Considering that for the deformations of the loop, $\delta x^\mu (\sigma = 0) = \delta x^\mu (\sigma = 2\pi) = 0$ and defining $F_{\mu\nu} \equiv \partial_\mu A_\nu - \partial_\nu A_\mu$ we get that the change of the holonomy due to a deformation of the loop is
\begin{equation}
\delta W = \left(\int_0^{2\pi}F_{\mu\nu}\frac{dx^\mu}{d\sigma}\delta x^\nu d\sigma \right)W.
\end{equation}

This shows that if the connection $A= A_\mu dx^\mu$ is flat, i.e., if $F_{\mu\nu} = 0$, then the holonomy is independent of the path over which it is calculated if these paths can be deformed into each other while their end-points remain fixed.

Now, since the set of loops is parameterized by $\tau \in [0,2\pi]$, we can write the expression above for the variation of the holonomy into a differential equation for $W$
\begin{equation}
\frac{dW}{d\tau} - \left(\int_0^{2\pi}F_{\mu\nu}\frac{\partial x^\mu}{\partial \sigma}\frac{\partial x^\nu}{\partial \tau}d\sigma \right)W = 0
\end{equation}
and the holonomy over $\gamma$ can be obtained by integrating this equation from $\tau = 0$ up to $\tau = 2\pi$.

Now if we consider the loop space $\mathcal{L}^{(1)}(\mathcal{M})$, the quantity 
\[
\mathcal{A}(\tau) \equiv \int_0^{2\pi}F_{\mu\nu}\frac{\partial x^\mu}{\partial \sigma}\frac{\partial x^\nu}{\partial \tau}d\sigma
\]
defines a connection evaluated at each loop in space-time which correspond to points in the loop space and by varrying these loops with continuous transformations we defined a path $\Gamma$ in $\mathcal{L}^{(1)}(\mathcal{M})$. 

The above equation which defines $W$ can be seen as the parallel transport equation in this loop space and therefore we consider the generalization of this equations and define the $2$-holonomy $V$ as satisfying
\begin{equation}
\label{app2}
\frac{dV}{d\tau} - \left(\int_0^{2\pi}G_{\mu\nu}\frac{\partial x^\mu}{\partial \sigma}\frac{\partial x^\nu}{\partial \tau}d\sigma \right)V = 0,
\end{equation}
where $G_{\mu\nu}$ is an anti-symmetric tensor.

Now, let us consider a closed $2$-dimensional surface $\Sigma$ which can be obtained by continuous deformations from another closed surface $\Sigma_0$ sharing a common point $x_R$ with each other. The $2$-holonomy $V$ can be calculated over $\Sigma$ by direct integration of (\ref{app2}) but can also be obtained as the result of the deformation of the surface $\Sigma_0$ into $\Sigma$, once $V$ over $\Sigma_0$ is known. 

In order to get $V$ following this second approach we need to find how it varies when we deform the closed surface. This is done in a similar way as it was done before to find $\delta W$: we start by considering the variation of the equation (\ref{app2}), then multiply the result by $V^{-1}$ and finally compute the variation of the term containing $G_{\mu\nu}$ explicitly and integrate the expression in $\tau$. This shall leave us with 
\begin{equation}
\delta V = \int_0^{2\pi}\int_0^{2\pi}\left(\partial_\lambda G_{\mu\nu}+\partial_\mu G_{\nu\lambda}+\partial_\nu G_{\lambda \mu}\right)\frac{\partial x^\mu}{\partial \sigma}\frac{\partial x^\nu}{\partial \tau}\delta x^\lambda d\sigma d\tau.
\end{equation}

From here we see that if $G$ is an exact $2$-form, then the $2$-holonomy is surface independent. 

Now, parameterizing this variation with $\zeta \in [0,2\pi]$, such that $\zeta = 0$ labels the surface $\Sigma_0$ and $\zeta = 2\pi$, the surface $\Sigma$, we get the differential equation
\begin{equation}
\frac{dV}{d\zeta} -\int_0^{2\pi}\int_0^{2\pi}\left(\partial_\lambda G_{\mu\nu}+\partial_\mu G_{\nu\lambda}+\partial_\nu G_{\lambda \mu}\right)\frac{\partial x^\mu}{\partial \sigma}\frac{\partial x^\nu}{\partial \tau}\frac{\partial x^\lambda}{\partial \zeta} d\sigma d\tau =0,
\end{equation} 
and integrating this equation in $\zeta$ will give us the desired $V$ on the surface $\Sigma$.

In the loop space $\mathcal{L}^{(2)}(\mathcal{M})$ the above equation defines the parallel transport through a $3$-dimensional volume in space-time, which corresponds to a path $\Gamma$ in the loop space, parameterized by $\zeta$. This leads us to introduce the $3$-holonomy $U$ defined by
\begin{equation}
\frac{dU}{d\zeta} -\int_0^{2\pi}\int_0^{2\pi}H_{\lambda \mu \nu}\frac{\partial x^\mu}{\partial \sigma}\frac{\partial x^\nu}{\partial \tau}\frac{\partial x^\lambda}{\partial \zeta} d\sigma d\tau =0,
\end{equation}
with $H_{\lambda \mu\nu}$ a completely anti-symmetric tensor.

\end{document}